\documentclass[aps,pra,twocolumn,superscriptaddress]{revtex4}

\usepackage{mathrsfs}
\usepackage{array}
\usepackage{graphicx}% Include figure files
\usepackage{dcolumn}% Align table columns on decimal point
\usepackage{bm}% bold math
\usepackage[title,titletoc,header]{appendix}
\usepackage{amssymb}
\usepackage{amsmath}
\usepackage{mathrsfs}
\usepackage{amsmath}
\usepackage{subfigure}
\usepackage{float}
\usepackage[title,titletoc,header]{appendix}
\usepackage{graphicx}
\usepackage{color}
\usepackage{amssymb}
\usepackage{graphicx}
\usepackage{verbatim}

\newcommand{\kk}{\mathbf{k}}
\newcommand{\qq}{\mathbf{q}}
\newcommand{\pp}{\mathbf{p}}
\usepackage{bbm}

\begin{document}

\title{Universal relations for spin-orbit coupled Fermi gas near an $s$-wave resonance}

\author{Pengfei Zhang}
\thanks{PengfeiZhang.physics@gmail.com}
\affiliation{Institute for Advanced Study, Tsinghua University, Beijing 100084, China}

\author{Ning Sun}
\thanks{ningsun.atom@gmail.com}
\affiliation{Institute for Advanced Study, Tsinghua University, Beijing 100084, China}

\date{\today}

\begin{abstract}  
The synthetic spin-orbit coupled quantum gases was widely studied both experimentally and theoretically in past decades. As shown in previous studies, this modification of single-body dispersion will in general couple different partial waves and thus distort the wave function of few-body bound states which determines the short-distance behavior of many-body wave function. In this work, we focus on the two-component Fermi gas with one-dimensional or three-dimensional spin-orbit coupling (SOC) near an $s$-wave resonance. Using the method of effective field theory and the operator product expansion, we derive universal relations for both systems, including the adiabatic theorem, Viral theorem, pressure relation, and obtain the momentum distribution matrix $\left<\psi^\dagger_a(\qq)\psi_b(\qq)\right>$ at large $\qq$ ($a,b$ are spin index). The momentum distribution matrix shows both spin-dependent and spatial anisotropic features. And the large momentum tail is modified at the next-to-the leading order thanks to the SOC. We also discuss the experimental implication of these results depending on the realization of the SOC. 
\end{abstract}

\maketitle
\section{Introduction} 
The ability for human to tune the interaction between neural atoms via Feshbach resonance \cite{Feshbach} largely enriches the systems realizable in cold-atom experiments. Among them, the strongly interacting ultracold quantum gases, which could be prepared by tuning the two-body scattering on resonance, attracts much attentions \cite{BECBCS,Braaten review}. Although it has been realized in experiments years ago, it is still hard to understand theoretically due to the lack of small parameters to control the perturbation theory. Nevertheless, the diluteness of the atomic gases indeed enables physicists to predict universal behaviors when perturbing the system with high frequency or large momentum sources. This set of beautiful relations, first proposed by Shina Tan in a series of papers for two-component Fermi gas near an $s$-wave resonance \cite{Tan}, is called the contact relation. It states that many observables such as the tail of the momentum distribution, the tail of radio-frequency(rf) spectral, the variation of energy along tuning the scattering length, are all determined by a single number $C$ called contact \cite{Tan,Braaten1,Leggett,Castin}. These relations has already been examined extensively in different experimental systems \cite{contact exp1,contact exp2,contact exp3}. The physical origin of contact relation is the fact that the many-body wave function should reduce to a two-body one when two atoms get close thanks to the hierarchy of the many-body scale (Fermi energy etc.) and the few-body scale (Van de Waals energy). Later, the contact relation is generalized to quantum gases on high partial-wave resonance \cite{Ueda,Zhenhua,Qi,D-wave} or with universal trimers \cite{three body,super,semi super}. Some of these new predictions have already been verified in experiments \cite{p-wave exp,three body exp1,three body exp2,three body exp3}. 

Instead of tuning the interaction strength, interesting physics can also emerge from altering the single-particle dispersion. As a concrete example, quantum gases with synthetic spin-orbit coupling (SOC) wherein the spatial motion of an atom is coupled to its hyperfine-state degree of freedom is another class of system that is widely studied both experimentally and theoretically in recent years \cite{SOC review3,Hui review1,Hui review2,soc exp1,soc exp2,soc exp3,magnetic soc,soc2d1,soc2d2}. Through controlling the atom-light interaction or gradient magnetic field, one can couple the neutral atoms to some external gauge potential that can be either abelian or not. Two important types of the SOC scheme are the one-dimensional (1D) scheme (the NIST type \cite{soc exp1}) where there is no $SO(3)$ rotational symmetry and the three-dimensional (3D) scheme where the total angular momentum is still conserved. In addition to its close relation to topological phases of insulators or superfluids \cite{topology}, the SOC also significantly affects the few-body physics when two-body interaction between atoms are considered and become important. Specifically, an isotropic 3D SOC will couple $s$ wave and $p$ wave for two atoms with zero total momentum, while in general case of finite total momentum or with the 1D SOC all partial waves would be mixed \cite{Xiaoling soc,Peng soc,Zhenhua soc}. As a result, modifications of Bethe-Peierls boundary condition is proposed. More interestingly, SOC affects the behavior of the universal bound states since it alters the single-particle density of states \cite{soc trimer}. 

Since SOC changes the behavior of the short-range wave function, it should also affect the universal behaviors of the system such as the tail of the momentum distribution. While the many-body physics with SOC has been broadly studied \cite{Hui review2}, the effect on the universal behaviors was only recently discussed for the momentum distribution averaged over direction with 3D SOC \cite{soc contact}. In a previous work it is shown that the $p$-wave boundary condition becomes non-universal with 3D SOC while $s$-wave boundary condition is always robust \cite{Xiaoling p-wave}. Hence we focus on the universal relations for two-component Fermi gas near an $s$-wave resonance in this work. Using the method of the effective field theory and the Operator Product Expansion (OPE) \cite{Peskin}, we derive the tail of the momentum distribution matrix $\left<\psi^\dagger_a(\qq)\psi_b(\qq)\right>$ where $a,b=\uparrow$ or $\downarrow$. Keeping the information about both the direction and the magnitude of the momentum, we find it both anisotropic and spin-dependent. The effect of SOC manifests itself in the modification to the next-to-leading order ($1/q^5$) in the large momentum tail, even more competitive to the contribution of effective range. Furthurmore, the experimental implications of these results are also discussed. We find it crucial whether the SOC is achieved in moving frame or lab frame. For the lab-frame SOC where the momentum distribution at large $q$ is directly modified in the lab frame, the effect can be captured by a time-of-flight measurement. While for the comoving-frame SOC it requires further consideration. 

This paper is organized as follows. First, we establish the effective field theory used in this paper. The validation of it is examined by comparing the phase shift of two-body scattering computed with this effective field-theory to the existing result in some special case. Next, we introduce new contacts for the SOC systems. In terms of them, some universal relations for the system including the adiabatic theorem, Viral theorem and the pressure relation are obtained. Then we study the momentum distribution matrix and discuss the SOC effect to the large momentum tail. The experimental implications with different SOC schemes are analysed subsequently. At last, there comes a brief summary and outlook.

\section{Effective field theory and two-body scattering}

\begin{figure*}[t]
\includegraphics[width=1.6\columnwidth]{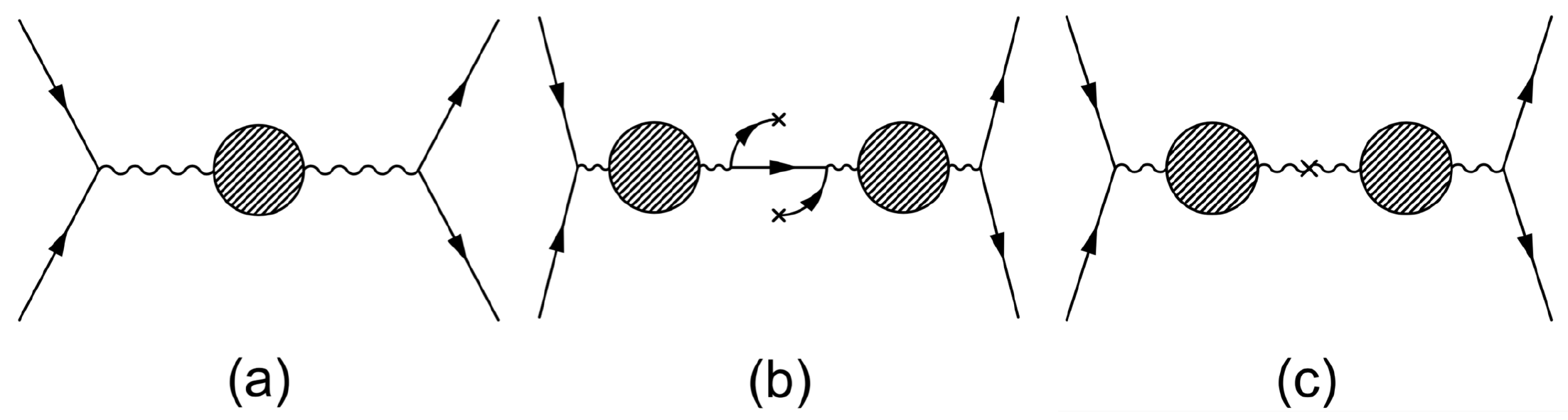}
\caption{Feynman diagrams for: (a) the $T$-matrix for the two-body scattering; (b) the matrix element of $\psi^{\dagger}_{\kk,a}\psi_{\kk,b}$ ; (c) the matrix element of dimer bilinears. In these diagrams, the wavy lines represent the propagators for the bare dimer fields, the solid lines represent the propagators for the fermion fields and the crosses represent the inserted operators.}
\label{dia1}
\end{figure*}

To study the universal physics for spin-orbit coupled system, we use the method of effective field theory. Here we take the effect of effective range into account and write down the following two-channel model:
\begin{align}
	\mathcal{L} =& \sum_{\kk}\psi_{\kk}^{\dagger}(i\partial_t\mathbbm{1}-h(\kk))\psi_{\kk}+d_{\kk}^{\dagger}(i\partial_t-k^2/4-\nu_0)d_{\kk} \notag\\
		& -\dfrac{g_0}{2\sqrt{V}}\sum_{\kk_1\kk_2}d_{\kk_1+\kk_2}^{\dagger}\psi_{\kk_1}\epsilon\psi_{\kk_2}+h.c. \label{S}
\end{align}
where we set $m=\hbar=1$ and $\psi=(\psi_{\uparrow}\; \psi_{\downarrow})^T$ hereinafter. $\psi_a$ is the annihilation operator of fermionic atoms with spin $a$ and $d$ is the annihilation operator for dimers. $\epsilon=i\sigma_y$ is the two-by-two antisymmetric matrix and $\mathbbm{1}$ denotes the two-by-two identity matrix. The first line of \eqref{S} gives the kinetic energy for free atoms and dimers with SOC, while the second line describes a conventional $s$-wave coupling written in the matrix form. $$\frac{1}{2}\psi \epsilon \psi=\frac{1}{2}(\psi_\uparrow\psi_\downarrow-\psi_\downarrow\psi_\uparrow)=\psi_\uparrow\psi_\downarrow.$$ Since the energy scale of the spin-orbit coupling is much smaller than that of the inter-atom potential, we assume the SOC affects only the single-atom dispersion $h(\kk)$ of the open channel. $h(\kk)$ for different cases is listed in Table. \ref{tab}.

\begin{table}[h]
\begin{tabular}{m{2.8cm} m{3cm}} 
 \hline\hline
  cases& $h(\kk)$\\ 
 \hline
 no SOC  & $\dfrac{k^2}{2}\mathbbm{1}$ \\ 
 1D SOC (NIST)  & $\dfrac{(\kk \mathbbm{1} +\lambda \sigma_z \hat{z})^2}{2}+\Omega \sigma_xå$ \\ 
 3D SOC& $\dfrac{(\kk \mathbbm{1} +\lambda \boldsymbol{\sigma})^2}{2}$   \\ 
  \hline\hline
\end{tabular}
 \caption{Single-particle Hamiltonian $h(\kk)$ used in the action \eqref{S} for different cases. Here $\sigma_i$ is the i-th Pauli matrix.}
 \label{tab}
\end{table}

Firstly, we focus on the two-body physics described by the $T$ matrix. As shown in the FIG. 1(a), the $T$ matrix is proportional to the exact propagator of the dimer field in vacuum. The Schwinger-Dyson equation for the exact propagator of the dimer field $G_{d}(t,\kk)=\left<\mathcal{T}d_{\kk}(t)d^\dagger_{\kk}(0)\right>$ is written as
\begin{align}
G^{-1}_d(E,\kk)=G^{-1}_{0,d}(E,\kk)-i\Pi_d(E,\kk),
\end{align}
where $G^{-1}_{0,d}(E,\kk)$ is the bare Green's function of dimers
\begin{align}
G_{0,d}(E,\kk)=\frac{i}{E^+-k^2/4-\nu_0},
\end{align}
and $i\Pi_d(E,\kk)$ is the self energy given by summing up the propagations of two fermions:
\begin{align}
i\Pi_d(E,\kk)
	&=-g_0^2\int^{|\qq|<\Lambda}\frac{d^3\qq dq_0}{(2\pi)^4}\frac{1}{2}\notag\\
	&\text{tr}\left[G^T(q_0,\mathbf{\frac{k}{2}+q})\epsilon G(E-q_0,\mathbf{\frac{k}{2}-q})\epsilon^\dagger\right].
\end{align}
The single-atom Green's function $G(\omega,\pp)$ is a two-by-two matrix
\begin{align}
\left[G^{-1}(\omega,\pp)\right]_{ab}=(-i)((\omega+i0^{+})\delta_{ab}-h(\pp)_{ab}).
\end{align}
and the integration over momentum in obtaining the self energy is truncated at a cutoff $\Lambda$.

Now the theory can be renormalized using few-body scattering parameters. It is done in the situation of no SOC, in which case the model reduces to a two-channel model with vanishing background scattering. The spirit here is the same as one usually do to renormalize the many-body physics ($k_F\neq0$) with a two-body ($k_F=0$) $T$ matrix. Aware of that $T\propto1/\cot(\delta)-i$, matching the phase shift expansion such that $k_r\cot\delta=-1/a+r_sk_r^2/2$, it leads to the following renormalization relations: 
\begin{align}
r_s&=-\frac{8\pi}{g_0^2},\\
\frac{1}{ a_s}&=\frac{-4\pi \nu_0}{g_0^2}+\frac{2\Lambda}{\pi}.\label{renormalization}
\end{align}
Note that although $r_s$ here is always negative, in general it can be positive near an $s$-wave resonance and the dimer field might be turned into a ghost field to cancel the minus sign \cite{Braaten2012}.

To demonstrate the effective field theory can also model the correct low-energy physics with the presence of SOC, we calculate the two-body phase shift for both 1D and 3D cases. It is straightforward to verify that the $\Lambda$ dependence can be canceled using the renormalization relation Eq. \eqref{renormalization} in both cases. For the special case of isotropic SOC with zero total momentum, the self-energy can be derived analytically,
\begin{align}
i\Pi_d(E,\mathbf{0})=g_0^2(\frac{i\Lambda}{2\pi^2}-\frac{1}{4\pi}\frac{E-\lambda^2}{\sqrt{E-2\lambda^2}}),
\end{align}
which yields the phase shift
\begin{align}
\cot\delta=(-\frac{1}{a_s}+\frac{E r_s}{2})\frac{\sqrt{E-2\lambda^2}}{E-\lambda^2}.
\end{align}
Here we assume that $E>2\lambda^2$ where $2\lambda^2$ is the threshold of two atoms with zero total momentum. It exactly reproduces the result in \cite{Xiaoling soc} when we set $r_s=0$ and shift the energy origin. From the perspective of solving Schr\"{o}dinger equation, it is directly related to the decoupling of different partial waves that we get such a simple analytical solution.

\section{Adiabatic theorem, Viral theorem and the pressure relation}

With no SOC, the physical parameters of the system are given by $a_s$ and $r_s$. It is known that two contacts can be defined to characterize the variation of energy along tuning each of them:
\begin{align}
C_a&\equiv\frac{dE}{d\left(-1/a_s\right)}=\left<\frac{dH}{d\left(-1/a_s\right)}\right>=\frac{g_0^2}{4\pi}\int d\mathbf{r}\left<d^\dagger d\right>,\\
C_r&\equiv\frac{dE}{dr_s}=\left<\frac{dH}{dr_s}\right>=\frac{g_0^2}{8\pi}\int d\mathbf{r}\left<d^\dagger (i\partial_t+\frac{\nabla^2}{4})d\right>.
\end{align}
Here $H$ is the Hamiltonian obtained via Legendre transformation of the action \eqref{S} and the Heisenberg equations for $d$ and $d^\dagger$ are used. These contacts are interesting for they govern other universal behaviors of a system, such as the Viral theorem, the pressure relation, the tail of the momentum distribution and the tail of rf spectral. In fact, the appearance of contact operator in the Viral theorem and the pressure relation is directly related to their definition. However, the behavior of the tail of the response functions is more non-trivial and needs further explanation which we defer to the next section. 

The operator form of the contacts directly suggests that the contacts merely count the number of dimers (weighted by the center of mass energy or not). Physical reason lies behind is that for a single dimer composed of two atoms, there are some relations between different observables. And the relations at short distance are robust against adding more particles. Moreover, it is the singularity of the two-body wave function that leads to the behavior of the tail of either momentum distribution or rf spectral. 

When SOC is present, there is a new parameter $\lambda$ for the 3D case, and an $\Omega$ in addition for the 1D case. Mimic how we deal with the scattering length $a_s$, naively one define $C_\lambda$ as
\begin{align}
C_\lambda\equiv\frac{dE}{d \lambda}&=\left<\frac{dH}{d\lambda}\right>,\label{Cldef}
\end{align} 
whose operator form is given by $\int d\mathbf{r}\left<\psi^\dagger\frac{dh}{d\lambda}\psi\right>$, and a similar definition works also for $C_\Omega$. Now the operator contains only single-atom operator which gives non-zero matrix elements in the single-atom sector. The momentum distribution under single-particle states (or direct product of that) is just a delta function and so is the rf spectral, hence $C_\lambda$ (and $C_\Omega$) will not contribute to a power-law tail in these measurables. Nevertheless, both $\lambda$ and $\Omega$ have non-zero energy scale, therefore they would appear in the Viral theorem and pressure relation. Both of the two relations can be derived straightforwardly using the Euler's homogeneous function theorem. For energy with a harmonic trap of frequency $\omega_{ha}$, $$E(\omega_{ha},a,r_s,\lambda,\Omega)=\omega_{ha} \mathcal{E}(\omega_{ha} a^2,\omega_{ha} r_s^2,\lambda^2/\omega_{ha},\Omega/\omega_{ha}),$$ and for the pressure, $$p(T,\mu,a,r_s,\lambda,\Omega)=T^{5/2}\phi(\mu/T,a^2 T,r_s^2 T,\lambda^2/T,\Omega/T).$$ Thus the Viral theorem for both cases are obtained as: 
\begin{align}
E &=2V_{ha}-\frac{1}{2a}C_a-\frac{r_s}{2}C_r+\frac\lambda 2C_\lambda\ \ \ \ \text{for 3D SOC}\\
E &=2V_{ha}-\frac{1}{2a}C_a-\frac{r_s}{2}C_r+\frac\lambda 2C_\lambda+\Omega C_\Omega\ \ \ \ \text{for 1D SOC}\end{align}
where the potential energy of the harmonic trap is given by $V_{ha}$. And the pressure relations are derived as:
\begin{align}
E &=\frac{3}{2}pV-\frac{1}{2a}C_a-\frac{r_s}{2}C_r+\frac\lambda 2C_\lambda\ \ \ \ \text{for 3D SOC}\\
E &=\frac{3}{2}pV-\frac{1}{2a}C_a-\frac{r_s}{2}C_r+\frac\lambda 2C_\lambda+\Omega C_\Omega\ \ \ \ \text{for 1D SOC}
\end{align}

Indeed there is an additional issue related to the details of experimental realization of the SOC action \eqref{S}. In experiments, SOC is usually achieved by shinning Raman laser beams to couple several atomic levels, the Hamiltonian of which differs from the SOC Hamiltonian in TABLE \ref{tab} by an unitary transformation. That is, $H_{\text{lab}}=U(\lambda)HU^\dagger(\lambda)$ with $U(\lambda)=\exp(i\psi^\dagger\sigma_z\psi \lambda z)$ for the 1D case, and similar for the 3D case. Hence the measurement of energy $ E$ in the lab frame should be actually using not the SOC Hamiltonian in TABLE \ref{tab} but the Hamiltonian in the lab frame. However, the argument that the tails of response response functions are independent of $C_\lambda$ and $C_\Omega$ is still correct, and these relations receive no further correction. For other experimental schemes to achieve the action \eqref{S} that do not acquire any frame transformation, such as using the gradient magnetic field \cite{magnetic soc}, there is no such issue. 

\section{anisotropic momentum distribution}
In this section we study another important universal relation, that is the tail of the momentum distribution. Here by saying the tail of the momentum distribution, we are in fact interested in a momentum regime where $|\mathbf q|^2$ is much larger than the Fermi energy while much smaller than the Van de Waals energy. When there is no SOC, it is known that the leading order of the momentum tail is proportional to $C_a/q^4$ and for systems with vanishing total momentum the sub-leading order, which contains the contribution of $C_r$, is proportional to $1/q^6$. 

Although none of new contacts defined via the adiabatic relation contributes to the momentum tail, as we discussed in the last section, one still expect different behavior due to the presence of SOC. Physically, we know that SOC should indeed make spin $\uparrow$ and $\downarrow$ different. Hence we consider the momentum distribution matrix $n_{ba}(\qq)=\left<\psi^\dagger_a(\qq)\psi_b(\qq)\right>$ and compute its matrix elements explicitly. $n_{ba}(\qq)$ is a $2\times2$ matrix in spin space. The diagonal elements give the expectation value of the number of atoms with either spin $\uparrow$ or spin $\downarrow$ and a certain momentum $\mathbf q$, which can be extracted by a time-of-flight measurement. The off-diagonal part indicates the mixing of different spins that can be measured in a similar way after an additional Rabi oscillation.

The technique we use to determine the large momentum dependence of $n_{ba}(\mathbf q)$ is OPE \cite{Peskin}. It is an operator relation that the product of two operators at small separation can be expanded in terms of the separation distance and operators, which can be interpreted as a Taylor expansion for the matrix elements of an operator.  
 \begin{align}
{O}_{i}\left(\mathbf{R}+\dfrac{\mathbf{r}}{2}\right){O}_{j}\left(\mathbf{R}-\dfrac{\mathbf{r}}{2}\right)
=\sum_{l} f_{ij,l}(\mathbf{r}){O}_{l}(\mathbf{R}). \label{ope}
 \end{align}
Here $O_i$ are local operators and $f_{ij,l}(\mathbf r)$ are expansion functions. Fourier transforming both sides of the equation above, it can also be equivalently performed in the momentum space. OPE is an ideal tool to explore short-range physics, that is $r_s\ll r\ll n^{-1/3}$ in a field theory context where $r_s$ is the effective range of the potential and $n$ is the average particle density. 

Following the conventional routine of deriving the tail of the momentum distribution, we match the matrix elements of $\hat{n}_{ba}(\qq)=\psi^\dagger_a(\qq)\psi_b(\qq)$ with contact operators in the two-body sector. Naively speaking, this means we consider the matrix elements of $\hat{n}_{ba}(\qq)$ between an incoming state of two atoms with spin $i_1$, $i_2$ and an outgoing state of two atoms with spin $i_3$, $i_4$. The total momentum and total energy of both incoming and outgoing states are given by $\kk$ and $E$ respectively. To calculate the matrix elements of contact operators, we consider the diagram, shown in FIG. \ref{dia1} (c), for a general operator $g_0^2\int d\mathbf r \ d^\dagger f(i\partial_t,-i\nabla) d$ where $f$ is an arbitrary function. As a result, 
\begin{align}
\langle\text{out}|g_0^2\int d\mathbf r \ d^\dagger &f(i\partial_t,-i\nabla)d|\text{in} \rangle=\notag\\&-\epsilon_{i_1,i_2}\epsilon^\dagger_{i_3,i_4}g_0^4 f(E,\kk)G_d^2(E,\kk)
\end{align}
Two special cases are $f=1$ and $f=E-\frac{k^2}{4}$. In that case the l.h.s. above reduces to $4\pi C_a$ or $8\pi C_r$ respectively. 

By choosing different $f$, one could match arbitrary function proportional to $-\epsilon_{i_1,i_2}\epsilon^\dagger_{i_3,i_4}g_0^4 G_d^2(E,\kk)$, which, as we will see, appears in the matrix elements of the the momentum distribution $\hat{n}(\qq)$. The diagrammatic representation of $\hat{n}(\qq)$ is shown in FIG. \ref{dia1} (b) and written as:
\begin{align}
\langle\text{out}|\hat{n}(\qq)|\text{in} &\rangle =-\epsilon_{i_1,i_2}\epsilon^\dagger_{i_3,i_4}g_0^4 G_d^2(E,\kk)\frac{(-i)^2}{(2\pi)^3}\int\frac{dq_0}{2\pi}\notag\\&G(q_0,\qq)\epsilon^\dagger G^T(E-q_0,\mathbf{k-q})\epsilon G(q_0,\qq).
\end{align}

In consideration of the momentum tail at large $q$, we do Taylor expansion of $n(\qq)$ matrix with respect to $1/q$. Since the $q^2$ term in the single-particle dispersion always dominates for either 1D or 3D SOC at large $q$, the leading order term is always $\propto 1/q^4$, and is spin-independent. The next-to-leading order $1/q^5$ is proportional to $\lambda^1$ or $\kk$. And it is the first term that is spin-dependent, thanks to SOC. 
\begin{align}
\text{For} &\  \text{1D SOC:}\notag\\
&n(\mathbf q)=\frac{1}{2\pi^2}\frac{C_a}{q^4}+\frac{1}{\pi^2} \frac{C_\mathbf k\cdot q}{q^6}-\frac{2\lambda}{\pi^2} \frac{\sigma_z\cdot q_z\ C_a}{q^6}+O(q^6)\label{1D}\\ 
\text{For} &\  \text{3D SOC:}\notag\\
&n(\mathbf q)=\frac{1}{2\pi^2}\frac{C_a}{q^4}+\frac{1}{\pi^2} \frac{C_\mathbf k\cdot q}{q^6}-\frac{2\lambda}{\pi^2} \frac{\sigma\cdot q\ C_a}{q^6}+O(q^6) \label{3D}
\end{align} 
Here we have defined an additional contact operator that do not appear in adiabatic relation, as we usually do when considering the next-to-leading order of the momentum tail:
\begin{align}
C_\mathbf k&\equiv\frac{g_0^2}{4\pi}\int d\mathbf{r}\left<d^\dagger (-i\nabla)d\right>.
\end{align}
Physically, $C_\mathbf k$ tells the total momentum of the dimer field. In many cases, it vanishes for an equilibrium system. Thus the next-to-leading order for such situation will be contributed directly from the SOC which makes different spin components inequivalent. With regard to experimental detection, such SOC effect should be observed in a time-of-flight measurement for the experimental scheme that directly realize SOC in the lab frame \cite{magnetic soc}. 

In contrast, for other experimental schemes, such as that mentioned in the last section which is different from \eqref{S} by an unitary transformation that boosts different spin components with different velocities, the detection of SOC effect needs further discussion. 

For 1D SOC case, The single-particle Hamiltonian in the lab frame is given by $$H(\kk)=\left(\frac{\kk^2}{2}+\Omega \sigma_+\exp(2i\lambda x)+\Omega \sigma_-\exp(-2i\lambda x)\right).$$ The first observation is that the SOC effect needs both $\lambda$ and $\Omega$ to be non-zero. For $\Omega=0$, the last two terms simply vanish; while for $\lambda=0$, the last two terms are equivalent to a Zeeman field and no net effect is induced for a dimer since the energy shifts cancel each other. 

As a result, if we shift the momentum and go back to the original frame by transforming $n_{\uparrow\uparrow}(\mathbf q)$ to $n_{\uparrow\uparrow}(\mathbf q-\lambda \hat z)$ and $n_{\downarrow\downarrow}(\mathbf q)$ to $n_{\downarrow\downarrow}(\mathbf q+\lambda \hat z)$, the result in \eqref{1D} is recast. Including also the spin-mixing (off-diagonal) terms, the momentum distribution matrix in the original frame is reckoned as:
\begin{align}
n(\mathbf q)_{\text{lab, spin-dependent}}=&(\frac{-8 q_z^2\lambda^2 \Omega C_a}{\pi^2q^{10}}+O(\frac{1}{q^9}))\sigma_x\notag\\&+(\frac{-8 q_z\lambda \Omega^2 C_a}{\pi^2q^{10}}+O(\frac{1}{q^{10}}))\sigma_z
\end{align}
And this result can be expected in regarding the two symmetries respected by the action \eqref{S}: i) $\psi\rightarrow \sigma_z\psi$ and $\Omega\rightarrow -\Omega$; ii) $\psi\rightarrow \sigma_x\psi$ and $\lambda\rightarrow -\lambda$. 

Now let's turn to the 3D case. Since the experimental realization utilizing Raman lasers involve more than two states \cite{3Dsoc}, a detailed analysis is much more complicated. Here we only ask whether the $n(\qq)_{\text{lab}}$ is given by $n_{\lambda=0}(\qq+\lambda \sigma)$. In fact, from straightforward expansion, we find:
\begin{align}
n(\qq)_{\text{lab}}=n_{\lambda=0}(\qq+\lambda \sigma)+O(1/q^6).
\end{align}
In a previous work \cite{range} $n_{\lambda=0}(\qq)$ was studied. As for the situation concerned here, the $n_{\lambda=0}(\qq+\lambda \sigma)$ is actually not well-defined to higher orders since $\sigma\cdot \mathbf k$ do not commute with $\sigma \cdot \mathbf q$. 

%By Taylor expansion, we find to the $1/q^6$ order, the matrix element $(2\pi)^3\langle\text{out}|\hat{n}(\qq)|\text{in}\rangle/(-\epsilon_{i_1,i_2}\epsilon^\dagger_{i_3,i_4}g_0^4 G_d^2(E,\kk))$ is given by:
%\begin{align}
%&\frac{1}{q^4}+\frac{2\mathbf q\cdot \mathbf k}{q^6}-\frac{4\lambda \mathbf q \cdot \sigma}{q^6}+\frac{2E-k^2}{q^6}+\frac{3(q\cdot \mathbf k)^2}{q^8}\notag\\&-\frac{12\lambda \mathbf q \cdot \sigma \mathbf q\cdot \mathbf k}{q^8}+\frac{2\lambda \kk\cdot \sigma}{q^6}+\frac{6\lambda^2}{q^6}+O(\frac{1}{q^7}).
%\end{align}
%It is straightforward to show that all $\lambda$ dependence can be written in terms of $n_{\lambda=0}(\qq+\lambda \sigma)$:
%\begin{align}
%\frac{1}{(\qq+\lambda \sigma)^4}&+\frac{2(\qq+\lambda \sigma)\cdot \mathbf k}{(\qq+\lambda \sigma)^6}\notag\\&+\frac{2E-k^2}{(\qq+\lambda \sigma)^6}+\frac{3(q\cdot \mathbf k)^2}{q^8}+O(\frac{1}{q^7}).
%\end{align}
%One could in principle define more contacts (where $f(E,\mathbf k)=k_ik_j$) to write out this universal tail which coincide with $n_{\lambda=0}(\qq+\lambda \sigma)$ for the two channel model for s-wave interaction taking the effective range into account \cite{range}. Indeed this form is itself not well-defined to the higher order since the $\sigma\cdot \mathbf k$ do not commute with $\sigma \cdot \mathbf q$. 

\section{Summary and Outlook}
In this paper, we study the two-component Fermi gas with SOC using effective field theory and discuss some universal behaviors near an $s$-wave resonance. Through the variation of energy along the SOC parameters, new contacts are defined and the Viral theorem and the pressure relation are derived in terms of them. Utilizing the technique of OPE, the tail of momentum distribution matrix are obtained. We find that it shows an anisotropic and spin-dependent behavior at the next-to-leading order thanks to SOC. The SOC manifest itself at the order $1/q^5$ of the momentum distribution matrix, even lower than the order of effective range $r_s$. In an experimental scheme realizing SOC directly in the lab frame, this can be observed through time-of-flight measurement. 

Besides, as is discussed in \cite{Xiaoling p-wave}, even on a $p$-wave resonance, one could not neglect the $s$-wave scattering. And the BP boundary condition is no longer robust. Resorting to the effective field theory approach, this means we need to add new bare parameters to a conventional two-channel $p$-wave model and new renormalization relations should be derived. Since the new parameters should be of the same scale as the $p$-wave effective range, they will contribute a non-zero momentum tail at $1/q^4$. Establishing an effective field theory model of this and deriving some universal relations on a $p$-wave resonance will be an interesting work which we defer to further study.

\textit{Acknowledgment.} We thank Xiaoling Cui for many valuable discussions on several important issues. We also thank Hui Zhai for helpful suggestions on the manuscript. 

\textit{Note added.} Upon finishing this work, we became to know a parallel paper on the similar topic \cite{Peng contact}. Our results are consistent.


\begin{thebibliography}{99}

\bibitem{Feshbach}
C. Chin, R. Grimm, P. Julienne, and E. Tiesinga, Rev. Mod. Phys. \textbf{82}, 1225 (2010).

\bibitem{BECBCS}
W. Zwerger, \textit{The BCS-BEC Crossover and the Unitary Fermi Gas}, Springer, 2011. 

\bibitem{Braaten review}
E. Braaten and H.-W. Hammer, Physics Reports, \textbf{428}: 259-390 (2006).

\bibitem{Tan}
S. Tan, Ann. Phys. (Amsterdam) \textbf{323}, 2952 (2008).

\bibitem{Braaten1}
E. Braaten and L. Platter, Phys. Rev. Lett. \textbf{100}, 205301 (2008).

\bibitem{Leggett}
S. Zhang and A. J. Leggett, Phys. Rev. A \textbf{79}, 023601 (2009).

\bibitem{Castin}
F. Werner, L. Tarruell, and Y. Castin, Eur. Phys. J. B \textbf{68}, 401 (2009).

\bibitem{contact exp1}
G. B. Partridge, K. E. Strecker, R. I. Kamar, M. W. Jack, and R. G. Hulet,
Phys. Rev. Lett. \textbf{95}, 020404 (2005).

\bibitem{contact exp2}
J. T. Stewart, J. P. Gaebler, T. E. Drake, and D. S. Jin, Phys. Rev. Lett. \textbf{104}, 235301 (2010).

\bibitem{contact exp3}
E. D. Kuhnle, H. Hu, X.-J. Liu, P. Dyke, M. Mark, P. D. Drummond, P. Hannaford, and C. J. Vale,
Phys. Rev. Lett. \textbf{105}, 070402 (2010).

\bibitem{Ueda}
S. M. Yoshida and M. Ueda, Phys. Rev. Lett. \textbf{115}, 135303 (2015).

\bibitem{Zhenhua}
 Z. Yu, J. H. Thywissen, and S. Zhang, Phys. Rev. Lett. \textbf{115}, 135304 (2015).

\bibitem{Qi}
M. He, S. Zhang, H. M. Chan, and Q. Zhou, Phys. Rev. Lett. \textbf{116}, 045301 (2016).

\bibitem{D-wave}
P. Zhang, S. Zhang and Z. Yu, Phys. Rev. A \textbf{95}, 043609 (2017).

\bibitem{three body}
E. Braaten, D. Kang and L. Platter, Phys. Rev. Lett. \textbf{106}, 153005 (2011).

\bibitem{super}
P. Zhang and Z. Yu, Phys. Rev. A \textbf{95}, 033611 (2017).

\bibitem{semi super}
P. Zhang and Z. Yu, Phys. Rev. A \textbf{96}, 030702 (2017).

\bibitem{p-wave exp}
C. Luciuk, S. Trotzky, S. Smale, Z. Yu, S, Zhang, and J.H. Thywissen, Nat. Phys. \textbf{12}, 1 (2016).

\bibitem{three body exp1}
D. H. Smith, E. Braaten, D. Kang, and L. Platter, Phys. Rev. Lett. \textbf{112}, 110402 (2014).

\bibitem{three body exp2}
M. Cetina, M. Jag, R. S. Lous, I. Fritsche, J. T. M. Walraven, R. Grimm, J. Levinsen, M. M. Parish, R. Schmidt, M. Knap and E. Demler, \textit{Science} \textbf{354} (6308), 96-99.

\bibitem{three body exp3}
R. J. Fletcher, R. Lopes, J. Man, N. Navon, R. P. Smith, M. W. Zwierlein, Z. Hadzibabic, \textit{Science} \textbf{355} (6323), 377-380.

\bibitem{SOC review3}
J. Dalibard, F. Gerbier, G. Juzeliūnas and P. Öhberg, Rev. Mod. Phys. \textbf{83}, 1523 (2011).

\bibitem{Hui review1}
H. Zhai, Int. J. Mod. Phys. B, 26, 1230001 (2012).

\bibitem{Hui review2}
H. Zhai, Rep. Prog. Phys. \textbf{78}, 026001 (2015).

\bibitem{soc exp1}
V. Galitski and I. B. Spielman, \textit{Nature} \textbf{494}, 49–54.

\bibitem{soc exp2}
P. Wang, Z.-Q. Yu, Z. Fu, J. Miao, L. Huang, S. Chai, H. Zhai, and J. Zhang,
Phys. Rev. Lett. \textbf{109}, 095301 (2012).

\bibitem{soc exp3}
L. W. Cheuk, A. T. Sommer, Z. Hadzibabic, T. Yefsah, W. S. Bakr, and M. W. Zwierlein
Phys. Rev. Lett. \textbf{109}, 095302 (2012). 

\bibitem{magnetic soc}
B. M. Anderson, I. B. Spielman and G. Juzeliunas, Phys. Rev. Lett. \textbf{111}, 125301 (2013).

\bibitem{soc2d1}
Z. Wu, L. Zhang, W. Sun, X.-T. Xu, B.-Z. Wang, S.-C. Ji, Y. Deng, S. Chen, X.-J. Liu and J.-W. Pan, \textit{Science} \textbf{354} (6308), 83-88.

\bibitem{soc2d2}
L. Huang, Z. Meng, P. Wang, P. Peng, S.-L. Zhang, L. Chen, D. Li, Q. Zhou and J. Zhang, \textit{Nature Physics} \textbf{12}, 540–544 (2016).

\bibitem{topology}
B. A. Bernevig and T. L. Hughes, \textit{Topological Insulators and Topological Superconductors}, Princeton University Press, 2013.

\bibitem{Xiaoling soc}
X. Cui, Phys. Rev. A \textbf{85}, 022705 (2012).

\bibitem{Peng soc}
P. Zhang, L. Zhang, and Y. J. Deng, Phys. Rev. A \text{86}, 053608 (2012).

\bibitem{Zhenhua soc}
Y. Wu and Z. Yu, Phys. Rev. A 87, 032703 (2013).

\bibitem{soc trimer}
Z.-Y. Shi, X. Cui, and H. Zhai, Phys. Rev. Lett. 112, 013201 (2014).

\bibitem{soc contact}
S.-G. Peng, C.-X. Zhang, S. Tan, K. Jiang, arXiv:1710.10579.

\bibitem{Xiaoling p-wave}
X. Cui, Phys. Rev. A 95, 030701(R) (2017).

\bibitem{Braaten2012}
E. Braaten, P. Hagen, H.-W. Hammer, and L. Platter
Phys. Rev. A {\bf86}, 012711 (2012). 

\bibitem{Peskin}
M.E. Peskin and D.V. Schroeder, \textit{An Introduction To
Quantum Field Theory}, Westview Press, Boulder,
(1995).

\bibitem{3Dsoc}
B. M. Anderson, G. Juzeliūnas, V. M. Galitski, and I. B. Spielman, Phys. Rev. Lett. 108, 235301 (2012).

\bibitem{range}
S. B. Emmons, D. Kang, and L. Platter, Phys. Rev. A \textbf{94}, 043615 (2016).

\bibitem{Peng contact}
J. Jie, R. Qi, and P. Zhang, to appear.

\end{thebibliography}
\end{document}